\documentclass[fleqn,12pt]{article}
\usepackage{epsfig}
\begin{document}
\textheight 22cm
\textwidth 15cm
\noindent
{\Large \bf Quasilinear analysis of the zonal flow back-reaction on ion-temperature-gradient mode turbulence}
\newline
\newline
Johan Anderson$^{a)}$\footnote{anderson.johan@gmail.com}, Jiquan Li$^{b)}$, Yasuaki Kishimoto$^{b)}$ and Eun-jin Kim$^{a)}$
\newline
$^{a)}$ Department of Applied Mathematics
\newline
Hicks Building, Hounsfield Road, University of Sheffield, Sheffield
\newline
S3 7RH, UK
\newline
$^{b)}$ Department of Fundamental Energy Science
\newline
Graduate School of Energy Science, Kyoto University, Gokasho, Uji, Kyoto 611-0011, Japan
\newline
\newline
\begin{abstract}
\noindent
There is strong evidence in favor for zonal flow suppression of the Ion-Temperature-Gradient (ITG) mode turbulence, specifically close to the linear stability threshold. The present letter attempts to analytically calculate the effects of zonal flow suppression of the ITG turbulence through deriving a modified dispersion relation including the back-reaction of the zonal flows on the ITG turbulence based on the quasilinear theory. The results are manifested in a reduction of the linear growth rate and an increase in the effective linear ITG threshold.
\end{abstract}
\newpage
\renewcommand{\thesection}{\Roman{section}}
\section{Introduction}
\indent
Anomalous transport remains one of the main concerns in magnetically confined plasmas. The anomalous transport in the core is commonly attributed to Ion-Temperature-Gradient (ITG) mode turbulence, however, other modes can not be totally excluded~\cite{a10}. An important critical concept is the transport regulation and transport barrier formation by zonal flows. The zonal flows are radially localized (radial wave number $q_x \neq 0$) and strongly sheared flows in the poloidal direction (poloidal and parallel wave number $q_y = 0$, $q_{\parallel} =0$)~\cite{a11}-~\cite{a14}. The large scale zonal flows are believed to be excited by the small scale fluctuating drift wave turbulence via Reynolds stress.

Several models for zonal flow generation from turbulence have been proposed, wave kinetic equation models, reductive perturbation methods and coherent mode coupling models~\cite{a14}. The coherent mode coupling model used in this particular study has been adopted and fitted to this field of study through the recent years~\cite{a14}-~\cite{a19}. It has been shown that the model is in good agreement with the corresponding model derived using the wave kinetic equation~\cite{a18} and the model has successfully contributed to increased understanding of the weak generation of zonal flows in recent measurements at the Columbia Linear Machine~\cite{a19}.

In this letter the effect of the back-reaction of zonal flows on the ITG mode turbulence is analytically investigated. Previous models on back-reaction of zonal flows on turbulence have mainly been concerned with linear estimates of shearing~\cite{a14},~\cite{a191}. Dynamical models for the coupled system of zonal flows and turbulence have also been proposed~\cite{a14},~\cite{a15}, however, no explicit estimate of the reduction of linear growth rate or up-shift of the critical gradient (e.g. $\nabla T_i$) due to zonal flow stabilization have been made~\cite{a20}. This letter is mainly concerned with estimating quantitative values of reduced linear growth rate and of the up-shift in the critical gradient. The inspiration for the research in the present letter is from the study of self-organization of plasma turbulence where zonal flows inhibits anomalous diffusion~\cite{a111} and the phenomenon of up-shift in the threshold value for the critical gradient or Dimits shift that has been observed in many non-linear simulations of ITG mode turbulence~\cite{a20}. A possible explanation for the Dimits shift is also recognized as one of the main yet unsolved physics issues for the understanding of zonal flows~\cite{a201}.

Based on the principle of quasilinear theory, the strategy of the letter is to first consider the zonal flow generation from the ITG background turbulence and then consider a small change in the drift wave perturbation calculated through the non-linearities. This small non-linear correction term is found by letting the zonal flow couple back to the side-bands resulting in a change in the background drift wave system. This back-reaction term can be interpreted as a change in the threshold of the critical gradient ($\eta_{ith \: NL}$) for the drift wave system. Moreover, the term changing the background is shown to cause a decreasing growth rate and an effective up-shift of the threshold. 

The letter is organized as follows. In Sec. II the model for the ITG mode is presented together with the model of zonal flow generation and back-reaction. In Sec. III the zonal flow saturation level is estimated and the results are presented in Sec IV. We provide a conclusion in Sec. VI.

\section{Governing equations and model}
The ITG mode turbulence is modeled using the continuity and temperature equation for the ions and considering the electrons to be Boltzmann distributed and the quasi neutrality is used to close the system~\cite{a21}-~\cite{a22} that has been successful in reproducing both experimental~\cite{a41} and non-linear gyro-kinetic results~\cite{a12}. The advanced fluid model is expected to give qualitatively and quantitatively accurate results in tokamak core plasmas or the flat density regime. For instance, this was shown in the plot of the slope of the ITG threshold as a function of $\epsilon_n$ in the flat density regime (see Figure 1 in Ref.~\cite{a42} or Figure 4 in Ref.~\cite{a43}) where this model recovers the full linear kinetic result within 5\%, far better than the prediction from an approximate kinetic model using a constant energy approximation in the magnetic drift. In the present work the effects of parallel ion motion, magnetic shear, trapped particles and finite beta on the ITG modes are neglected. It has been found earlier that the effect of parallel ion motion on the ITG mode is rather weak, however, it has recently been shown that magnetic shear and parallel ion motion may modify the zonal flow generation significantly~\cite{a23}. The zonal flow generation from Trapped Electron Modes~\cite{a24} and the effects of parallel ion motion~\cite{a19} on zonal flow generation have been treated earlier.

The continuity and temperature equations are,{\bf
\begin{eqnarray}
\frac{\partial \tilde{n}}{\partial t} - \left(\frac{\partial}{\partial t} - \alpha_i \frac{\partial}{\partial y}\right)\nabla^2_{\perp} \tilde{\phi} + \frac{\partial \tilde{\phi}}{\partial y} - \epsilon_n g_i \frac{\partial}{\partial y} \left(\tilde{\phi} + \tau \left(\tilde{n} + \tilde{T}_i \right) \right) = \nonumber \\
- \left[\tilde{\phi},\tilde{n} \right] + \left[\tilde{\phi}, \nabla^2_{\perp} \tilde{\phi} \right] + \tau \left[\tilde{\phi}, \nabla^2_{\perp} \left( \tilde{n} + \tilde{T}_i\right) \right] \\
\frac{\partial \tilde{T}_i}{\partial t} - \frac{5}{3} \tau \epsilon_n g_i \frac{\partial \tilde{T}_i}{\partial y} + \left( \eta_i - \frac{2}{3}\right)\frac{\partial \tilde{\phi}}{\partial y} - \frac{2}{3} \frac{\partial \tilde{n}}{\partial t} = \nonumber \\
- \left[\tilde{\phi},\tilde{T}_i \right] + \frac{2}{3} \left[\tilde{\phi},\tilde{n} \right].
\end{eqnarray}}
The system is closed by using quasi-neutrality with Boltzmann distributed electrons. Here $\left[ A ,B \right] = (\partial A/\partial x) (\partial B/\partial y) - (\partial A/\partial y) (\partial B/\partial x)$ is the Poisson bracket. With the additional definitions $\tilde{n} = \delta n / n_0$, $\tilde{\phi} = e \delta \phi /T_e$, $\tilde{T}_i = \delta T_i / T_{i0}$ as the normalized ion particle density, the electrostatic potential and the ion temperature, respectively. In the forthcoming equations $\tau = T_i/T_e$, $\vec{v}_{\star} = \rho_s c_s \vec{y}/L_n $, $\rho_s = c_s/\Omega_{ci}$ where $c_s=\sqrt{T_e/m_i}$, $\Omega_{ci} = eB/m_i c$. We also define $L_f = - \left( d ln f / dr\right)^{-1}$, $\eta_i = L_n / L_{T_i}$, $\epsilon_n = 2 L_n / R$ where $R$ is the major radius and $\alpha_i = \tau \left( 1 + \eta_i\right)$. The perturbed variables are normalized with the additional definitions $\tilde{n} = (L_n/\rho_s) \delta n / n_0$, $\tilde{\phi} = (L_n/\rho_s e) \delta \phi /T_e$, $\tilde{T}_i = (L_n/\rho_s) \delta T_i / T_{i0}$ as the normalized ion particle density, the electrostatic potential and the ion temperature, respectively. The perpendicular length scale and time are normalized to $\rho_s$ and $L_n/c_s$, respectively. The geometrical quantities are calculated in the strong ballooning limit ($\theta = 0 $, $g_i \left(\theta = 0, \kappa \right) = 1/\kappa$  where $g_i \left( \theta \right)$ is defined by $\omega_D \left( \theta \right) = \omega_{\star} \epsilon_n g_i \left(\theta \right)$)~\cite{a25}-~\cite{a26}.
The approximate linear solutions to Eqs 1 and 2 are,
\begin{eqnarray}
\omega_{r DW} & = & \frac{k_y}{2\left( 1 + k_{\perp}^2\right)} \left( 1 - \left(1 + \frac{10\tau}{3} \right) \epsilon_n g_i - k_{\perp}^2 \left( \alpha_i + \frac{5}{3} \tau \epsilon_n g_i \right)\right),  \\
\gamma_{DW} & = & \frac{k_y}{1 + k_{\perp}^2} \sqrt{\tau \epsilon_n g_i \left( \eta_i - \eta_{i th}\right)}.
\end{eqnarray}
where $\omega = \omega_r + i \gamma$ and 
\begin{eqnarray}
\eta_{i th} \approx \frac{2}{3} - \frac{1}{2 \tau} + \frac{1}{4 \tau \epsilon_n g_i} + \epsilon_n g_i \left( \frac{1}{4 \tau} + \frac{10}{9 \tau}\right).
\end{eqnarray}
Finite Larmor Radius (FLR) effects in the $\eta_{i th}$ are neglected, here, however, in the remaining part of the letter the linear state is solved numerically and FLR effects are included.

Following the same procedure as in Refs~\cite{a15},~\cite{a18}-~\cite{a19} by choosing a monochromatic pump wave that generates a secondary convective cell. This gives a problem with four coupled waves where the zonal flow seed may be unstable and grow exponentially. Here, the sidebands may couple to the pump wave to give the zonal flow component. No Boltzmann distribution for electrons is given to derive the third order dispersion relation for the zonal flow component~\cite{a18},
\begin{eqnarray}
(\Omega + i \mu q_x^2)((1+k_{\perp}^2)\Omega^2 - \alpha^2) & = & K_0 [ q_x(1+k_{\perp}^2) \Omega - 2 k_x \alpha] |\phi_{DW}|^2, \\
K_0 & = & 2 \beta k_y^2 q_x k_{\perp -}, \\
k_{\perp -} & = & k_x^2 + k_y^2 - q_x^2, \\
\beta & = & 1 + \tau + \tau \delta, \\
\delta(\omega) & = & \frac{(\eta_i - \frac{2}{3})k_y + \frac{2}{3}\omega}{\omega + \frac{5}{3}\tau \epsilon_n g_i k_y}, \\
 \alpha & = & (1+k_{\perp}^2)\omega - k_y + \epsilon_n g_i k_y \beta.
\end{eqnarray}
Here, $\Omega$ is the zonal flow real frequency and growth rate and $q_x$ is the zonal flow wavenumber. In this letter, the third order dispersion relation is the starting point for the calculating the back-reaction of zonal flows on the ITG mode turbulence. 

To calculate the back-reaction, we start by calculating a small perturbation to the linear state using Eq. 1 and utilizing the dominant nonlinear coupling of the sidebands to the zonal flow to back-react on the turbulence,
This can be written as,
\begin{eqnarray}
\tilde{T}_{i \: DW} = L_0 \tilde{\phi} + \frac{i q_x^3 \beta}{\epsilon_n g_i \tau}(\tilde{\phi}_{+} \tilde{\phi}_{ZF}^* - \tilde{\phi}_{-}^* \tilde{\phi}_{ZF}).
\end{eqnarray}
Here $L_0$ is the linear part of the solution,
\begin{eqnarray}
L_0 & = & - \frac{\omega + (\omega + \alpha_i k_y) k^2_{\perp}- (1-(1+\tau)\epsilon_n g_i)k_y}{\epsilon_n g_i \tau k_y}. 
\end{eqnarray}

 It is also assumed that in calculating the non-linearities the linear relationship between the electric potential perturbation ($\tilde{\phi}$) and the ion temperature perturbation ($\tilde{T_i}$) can be used. Here, it is important to note that the positive and negative sideband couples through the non-linear terms only.

Next, the sideband relationships are considered. The equations for the positive and negative sidebands can be written,
\begin{eqnarray}
\tilde{\phi}_{+} & = & i \beta q_x k_y k_{\perp -} \frac{\tilde{\phi}_{DW} \tilde{\phi}_{ZF}}{(1+k_{\perp}^2)\Omega + \alpha}, \\ 
\tilde{\phi}_{-} & = & - i \beta q_x k_y k_{\perp -} \frac{\tilde{\phi}^{*}_{DW} \tilde{\phi}_{ZF}}{(1+k_{\perp}^2)\Omega - \alpha}.
\end{eqnarray}
The effects of the ion temperature non-linearity ($[\phi, T_i]$) are neglected, if this is taken into account the sidebands become coupled. Moreover, if the full system is studied including the ion temperature non-linearity the result found here is recovered in an expansion in the zonal flow wave number ($q_x$) and keeping the leading order term. The new real frequency and growth rate found from the perturbed linear state is renamed ($\bar{\omega}$). Utilizing the sideband Eq.s 14 and 15 into the perturbation to the linear state Eq. 12 the modified temperature perturbation can be found,
\begin{eqnarray}
\tilde{T}_{i DW} & = & L_0 \tilde{\phi}_{DW} \nonumber \\
& - &  K_1 (\frac{1}{(1+k_{\perp}^2)\Omega + \alpha} - \frac{1}{(1+k_{\perp}^2)\Omega^{*} - \alpha^{*}})|\tilde{\phi}_{ZF}|^2 \tilde{\phi}_{DW}, \\
& = & L_0 \tilde{\phi}_{DW} - NL |\tilde{\phi}_{ZF}|^2 \tilde{\phi}_{DW}, \\
K_1 & = & \beta^2 q^4_x k^2_y k_{\perp -}.
\end{eqnarray}
Eq. 17 is now used together with the linear part of Eq. 2,
\begin{eqnarray}
\tilde{T}_{i DW} = \frac{(\eta_i - \frac{2}{3})k_y + \frac{2}{3}\omega}{\omega + \frac{5}{3} \tau \epsilon_n g_i k_y} \tilde{\phi}_{DW}.
\end{eqnarray}
Combining Eq.s 17 and 19 a modified second order algebraic equation for the linear state modified by the zonal flow can be found,
\begin{eqnarray}
(1+k_{\perp}^2) \bar{\omega}^2 + (-(1-(1+\frac{10 \tau}{3})\epsilon_n g_i & - & \nonumber \\
 - k^2_{\perp}(\alpha_i + \frac{5}{3}\tau \epsilon_n g_i)) + \frac{NL}{k_y} |\tilde{\phi}_{ZF}|^2 )k_y \bar{\omega} & + & \nonumber \\
(\eta_i - \frac{7}{3} + \frac{5}{3}(1+\tau)\epsilon_n g_i + \frac{5}{3} \alpha_i k^2_{\perp} + \frac{5 NL}{3 k^2_y} |\tilde{\phi}_{ZF}|^2) k^2_y \tau \epsilon_n g_i & = & 0.
\end{eqnarray}
This equation describes a linear drift wave state modified by the effects of zonal flows. Moreover, it is seen that the zonal flow saturation level can directly modify the ITG mode stability threshold that may correspond to the so-called Dimits shift, observed in numerous non-linear transport simulations~\cite{a20}. 

\section{Saturation levels}
To quantify the effects of the zonal flow suppression an estimate of the zonal flow saturation level is needed. The zonal flow saturation may be subject to several nonlinear mechanisms in addition to the neoclassic collisional damping. In a collisionless plasma, possible nonlinear processes to limit the zonal flow instability include the spectral modulation of the ambient turbulence due to the flow, tertiary instability like the familiar Kelvin-Helmholtz(KH) mode, the nonlinear wave-packet scattering and/or trapping~\cite{a14},~\cite{a30}. Here to sidestep the complex, specific nonlinear process, an approximate method similar to the mixing length theory for estimating the drift wave turbulence saturation level, which involves the spectral modulation process of the turbulence, is applied to roughly evaluate the zonal flow level and the parametric dependence on the turbulent pump fluctuation. In previous research the drift wave saturation level has been estimated through mixing length theory,
\begin{eqnarray}
\tilde{\phi}_{DW} \sim  \frac{1}{k_x L_{T_i}}. 
\end{eqnarray} 
This estimate gives a relation between the drift wave electric potential perturbation and the temperature gradient length scale. It is assumed that the main source of free energy is the temperature gradient. The zonal flow saturation level can be estimated through a balance of the vorticity equation. The mode coupling saturation level for drift waves is found by balancing the linear drift wave growth rate with the main ($\vec{E} \times \vec{B}$ convection) non-linearity. In analogy, the zonal flow saturation level is estimated by balancing the linear growth rate with the Reynolds stress term, 
\begin{eqnarray}
\frac{\partial \nabla^2_{\perp}}{\partial t} \tilde{\phi}_{ZF} & \sim  & \nabla^2_{\perp} \langle \frac{\partial \tilde{\phi}_{DW}}{\partial x} \frac{\partial \tilde{\phi}_{\pm}}{\partial y} \rangle 
\end{eqnarray}
It is argued that the saturation of the zonal flow instability occurs when the sidebands become comparable to the pump wave~\cite{a51}, i.e., $|\tilde{\phi}_{\pm }| \approx | \tilde {\phi}_{DW}|$. Hence,  
\begin{eqnarray}
\tilde{\phi}_{ZF} \gamma_{ZF} & \sim  & \frac{k_x^3 k_y}{q_x^2} |\tilde{\phi}_{DW}|^2 = \lambda \frac{k_y k_x} {q_x^2} \frac{1}{L^2_{T_i}} 
\end{eqnarray}
Here $\lambda $ is a proportionality constant. From Eq. 23 it is possible to calculate the zonal flow saturation level. In the last equality Eq. 21 has been used. This saturation level gives some interesting insights, the zonal flow saturation level is increased by a streamer like source drift wave (large $k_y$ and small $k_x$) similar results were found in Ref.~\cite{a31} and it is similar to previous results found in gyrokinetic theory~\cite{a27}-~\cite{a28}. 

To corroborate the analytical estimation of the zonal flow saturation level numerical calculations based on the Hasegawa-Mima turbulence system, which is extensively employed to investigate the zonal flow generation~\cite{a14}-~\cite{a51} and also saturation level~\cite{a14},~\cite{a51}, have been performed for the equation~\cite{a16}-~\cite{a51} 
\begin{eqnarray}
(1-\delta -\nabla_{\perp}^2)\frac{\partial \phi}{\partial t} = \frac{\partial \phi}{\partial y} - \frac{\partial {\phi}_{ZF}}{\partial x} \frac{\partial \phi}{\partial y}  +  [\phi,\nabla_{\perp}^2 \phi],
\end{eqnarray} 
using an assumed pump wave of the form
\begin{eqnarray}
\phi_p = A_p \sin (k_x x) \cos (k_y y).
\end{eqnarray} 
Here $\delta =1 (0)$ for the zonal flow (finite $ k_y$ ) components, respectively. In the calculation, pump energy decreases as the zonal flow instability is excited while the total energy is conserved. We change the total pump energy under fixing $k_x$ ($=0.125$ or $0.25$) and $ k_y = 1$. In the Hasegawa-Mima system a non-linear saturation of zonal flows is present and a good agreement with the proportionality $\tilde{\phi}_{ZF} \gamma_{ZF} \sim |\tilde{\phi}_{DW}|^2 $ in (Eq. 23) is found, as shown in Figure 1. Hence, although Eq.(23) is not rigorously justified analytically, the numerical calculations show that the relation between the zonal flow saturation level and the pump ITG fluctuation is qualitatively correct and is quantitatively represented by the proportionality constant, which may involve the spectral relation and other physical processes. 

\begin{figure}
  \includegraphics[height=.3\textheight]{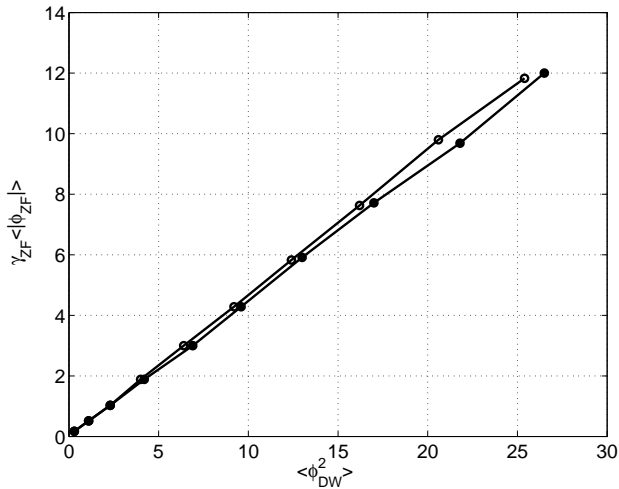}
  \caption{The zonal flow saturation level as a function of drift wave pump energy. $k_x=0.125$ (open circles) or $k_x=0.25$ (closed circles), $k_y=1.0$, $q_x=0.75$.}
\end{figure}

\section{Result and discussion}
The algebraic dispersion relation (Eq. 20), describing the effects of zonal flow back-reaction on the background turbulence, is solved in two cases using the derived saturation levels for drift waves and zonal flows. 

In Figure 2 the drift wave growth rate as a function of normalized temperature gradient ($\eta_i$) is shown with zonal flow wave number ($q_x$) as a parameter for $\tau = 1$, $k_x = k_y = 0.3$, $\epsilon_n = 1.0$, $\mu = 0$. The results are displayed for $q_x = 0.0$ (squares) and $q_x = 0.8$ (plus). Figure 2 shows that the zonal flow has a significant suppression effect on the drift wave growth rates close to the threshold, the growth rates are reduced by  $\geq 50$\% in the range $1.8 < \eta_i < 2.3$. The critical gradient is also modestly increased by the zonal flow suppression.

\begin{figure}
  \includegraphics[height=.3\textheight]{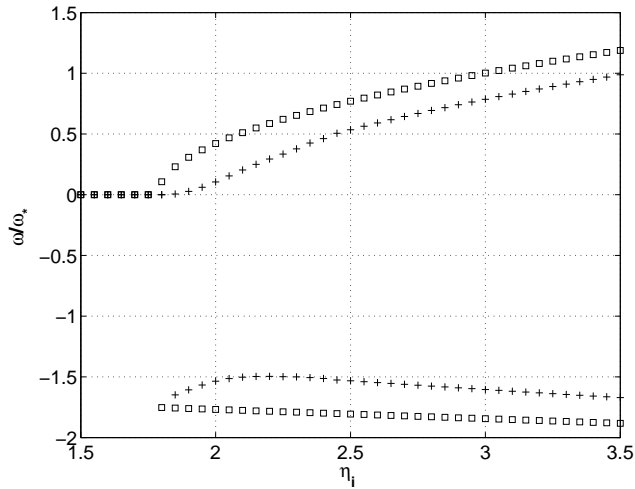}
  \caption{The drift wave growth rate and real frequency as a function of normalized temperature gradient ($\eta_i$)is shown with zonal flow wave number ($q_x$) as a parameter for $\tau = 1$, $k_x = k_y = 0.3$, $\epsilon_n = 1.0$, $\mu = 0$. The results are displayed for $q_x = 0.0$ (squares) and $q_x = 0.8$ (plus).}
\end{figure}

Figure 3 illustrates the effect of zonal flow back-reaction on the critical gradient ($\eta_{i th}$) in the $\epsilon_n$ - $\eta_i$ plane for short wave length zonal flows ($q_x = 0.8$). The parameters are $\tau = 1$, $k_x = k_y = 0.3$, $\mu = 0$, $q_x = 0.0$ (squares) and $q_x = 0.8$ (asterisk). It is found that the critical gradient is significantly increased for peaked density profiles (small $\epsilon_n$) relevant for edge plasmas. For both cases the FLR effects tend to decrease the stability threshold for peaked density profiles whereas the non-linear contribution tends to increase the threshold.

\begin{figure}
  \includegraphics[height=.3\textheight]{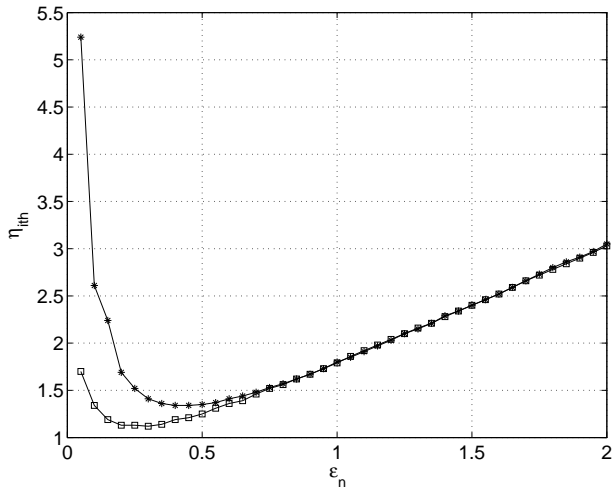}
  \caption{The effect of zonal flow back-reaction on the critical gradient ($\eta_{i th}$) in the $\epsilon_n$ - $\eta_i$ plane for short wave length zonal flows ($q_x = 0.8$) is shown. The parameters are $\tau = 1$, $k_x = k_y = 0.3$, $\mu = 0$, $q_x = 0.0$ (squares) and $q_x = 0.8$ (asterisk).}
\end{figure}

At present, the temperature non-linearity has been neglected for simplicity, if it is incorporated more involved sideband expressions will be found where the sidebands couple, however, the effect will be a factor $q^4_x$ smaller than the terms included here.

The effect of zonal flow suppression is generally manifested in a decreasing linear growth rate and/or the up-shift of the critical gradient (stability threshold). There are other possible stabilization mechanisms and it is interesting to note that a change in the spectrum due to the back-reaction of the zonal flow may also change the diffusion, moreover, it has been shown that the spectrum as well as the drift wave growth rate are quite prone to plasma shaping effects~\cite{a18}-~\cite{a19} which is out of the scope of the present study.

\section{Conclusion}
In summary, the present study aims to clarify the explicit effects of zonal flow suppression on drift waves based on a quasilinear theory. The short wave length zonal flow suppression of drift wave turbulence is significant in particular close to the stability threshold ($\eta_{ith}$). In the regime above the linear threshold the drift wave growth rate is significantly reduced indicating a significantly reduced transport. To explain the total quenching of turbulence in the wider Dimits shift region a more complete model is needed including more physics as well as plasma shaping effects. Numerical calculations of realistic ITG turbulence with focus on the zonal flow saturation level is also planned. These will be treated in future papers.

\section{Acknowledgments}
This research was partly supported by the Grant-in-Aid from Japan Society for the Promotion of Science (No. 18340186 and 19560828) and Engineering and Physical Sciences Research Council (EPSRC) (No. EP/D064317/1).

\end{document}